\documentclass{llncs}
\usepackage[utf8]{inputenc}
\usepackage{verbatim}
\usepackage{graphicx}
\usepackage{listings}
\usepackage{color}
\usepackage{url}
\usepackage{subcaption}
\title{The Declaratron, semantic specification for scientific computation using MathML}
\author{Dave Murray-Rust\inst{1} \and Peter Murray-Rust\inst{2}}
\institute{\email{d.murray-rust@ed.ac.uk}, Department of Informatics, University of Edinburgh \and \email{pm286@cam.ac.uk}, Department of Chemistry, University of Cambridge}
\date{March 2013}

\newcommand{\fig}[3][0.9]{
\begin{figure}[tp]
\begin{center}
\includegraphics[width=#1\textwidth]{#2}
\caption{#3}
\label{fig:#2}
\end{center}
\end{figure}
}

\begin{document}
\pagenumbering{arabic}
\pagestyle{headings}
\maketitle
\begin{abstract}
We introduce the Declaratron, a system which takes a declarative approach to specifying mathematically based scientific computation. This uses displayable mathematical notation (Content MathML) and is both executable and semantically well defined. We combine domain specific representations of physical science (e.g. CML, Chemical Markup Language), MathML formulae and computational specifications (DeXML) to create executable documents which include scientific data and mathematical formulae. These documents preserve the provenance of the data used, and build tight semantic links between components of mathematical formulae and domain objects---in effect grounding the mathematical semantics in the scientific domain. The Declaratron takes these specifications and i) carries out entity resolution and decoration to prepare for computation ii) uses a MathML execution engine to run calculations over the revised tree iii) outputs domain objects and the complete document to give both results and an encapsulated history of the computation. A short description of a case study is given to illustrate how the system can be used. Many scientific problems require frequent change of the mathematical functional form and the Declaratron provides this without requiring changes to code. Additionally, it supports reproducible science, machine indexing and semantic search of computations, makes implicit assumptions visible, and separates domain knowledge from computational techniques. We believe that the Declaratron could replace much conventional procedural code in science.
\end{abstract}

\setcounter{footnote}{0}


\section{Introduction}

\emph{This manuscript is offered as a Work-in-Progress with the primary motivation of bridging the current gap between mathematics markup communities and physical scientists. The Declaratron is a  system accessible to both communities and designed for collaborative working.}

Computational physical science is now recognised as a key part of modern science \cite{hey2009fourth}. However, there is heavy use of 40-year-old FORTRAN codes, which makes it extremely hard to reformulate and recalculate problems on-the-fly, and to reproduce results \cite{stodden2012reproducible,leveque2012reproducible}. Problems include undocumented program ``tweaks'', semantic ambiguities (e.g. units of measurement) and unreliable parameter values (e.g. out-of-date constants). 
The increasing importance and usage of formal semantics---highlighted in \cite{murray2011semantic,murray2012semantic}---leads us to propose a system where domain spemantics is made explicit, to the point where a software engineer without domain (in this case chemical) knowledge could implement and validate a processing engine correctly. Our Declaratron system uses \emph{datuments}---a document mixing data with mathematical relationships and presentation \cite{murray2007mathematics}---to take a declarative rather than procedural approach to scientific computation.
We make use of MathML\cite{ausbrooks2003mathematical} and Chemical Markup Language\footnote{the example uses CML, but the approach is applicable to any ML which manages numbers or geometry (e.g. GeographyML)} (CML)\cite{murray1999chemical,murray2003chemical} for representation of data and computation. 

This is demonstrated through an example, ``molecular forcefields" which computes the approximate energies of molecules, and is an extremely common task in computational chemistry. It is abstractable to four components:
\begin{enumerate}
\item \textbf{the scientific domain-objects} to be computed (molecules, atoms, their Cartesian coordinates and notional “bonds” between certain pairs).
\item \textbf{the functional form (FF)} of the energy function, which, at its simplest can be approximated by  Hooke’s Law, but there are hundreds of variants, often with many terms. For example, the GULP \cite{gale1997gulp} program's manual \cite[pp22-27]{gulpManual} gives an excellent impression of the variety. 
\item \textbf{the parameters} relating a given molecule to any given FF. These change fairly frequently as the science develops.
\item \textbf{the problem to be computed}, which can include single-point calculation, optimisation of energy, calculation of second derivatives (vibrational frequencies), dynamical calculations for integrating Newton’s laws (e.g. Verlet, \cite{grubmuller1991generalized}) into trajectories.
\end{enumerate}

To generalise, we have a formula to be used (item 2), some data to use in the computation (items 1 and 3) and a specification for the kind of computation to be done (item 4). Items 1 and 3 can also be reduced to a system of tested independent modules ("black-boxes"); in this case, JUMBO \cite{zhang2004jumbo} provides this for chemistry, with code that represents atoms and molecules, and can calculate basic properties such as bond lengths and angles.

The simplest forcefield is a quadratic equation, describing the approximate force between each pair of bonded atoms:
$$E=\sum_{bonds}{a(l-l_0)^2}$$

However $E, a, l$ and $l_0$ are semantically unbound---they are symbols unrelated to the physical world. In order to perform a calculation, we need to know that $a$ is a constant, which is different for any given atom pair, $l_0$ is the ideal interatomic distance, and $l$ is the actual interatomic distance (which can vary in an optimisation or trajectory calculation). These relationships generally need to be inferred from context, unless specifically stated in the surrounding text.
In order to use this formula in a computation, we need to, at a minimum, i) know that $E$ is an energy to calculate; ii) know that the summation is over the set of bonds in a given molecule (and have some idea what a bond means); iii) realise that there is an invisible subscript on the $a$, and it is different for each bond type iv) know that $l$ should be calculated as a 3D distance between the two atoms in a particular bond v) know that $l_0$ has another invisible subscript, and needs to be looked up for a given bond. And, given all of that, it is still not clear where to get the data to compute over, let alone what the units are, or the provenance of the data.

This issue becomes becomes more acute when we consider e.g. the forcefield equation used in AMBER---a popular program for calculating and optimising molecular forcefields---shown in Figure \ref{fig:amber}. Leaving aside the typo (there is a missing ')'), problems include: i) what are the precise elements of the sets (bonds, angles, dihedrals, nonbij)? ii) "dihedrals" should be  a double sum including Fourier terms (n) iii) the electrostatic section ($\sum_{nonbij}$) is missing a constant $4 \pi \epsilon_0$. Additionally, the parts of the equation are named differently by different people: dihedrals can be called torsions, nonbij means non-bonded, but this part of the equation is often called "electrostatics". This is not a carefully chosen example of poor specficiation; rather it is an illustration of common current practice. 

\fig[0.6]{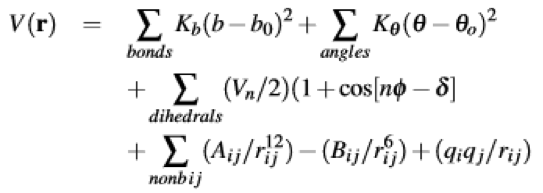}{The AMBER molecular forcefield equation, taken from the AMBER manual \cite[p19]{AMBER11}.} 

\subsection{Goals}

The Declaratron uses MathML to allow users to clearly and explicitly encode all of the necessary structures for scientific computation, in a domain-independent, standards compliant, machine readable manner. By separating domain knowledge from computation, we hope to allow software engineers with little or no domain knowledge to construct and validate the computational infrastructure, while domain experts with less computational knowledge can create the links between data, formulae and computational specification. The requirement for explicit semantic bindings between MathML statements and other (scientific) statements creates a more transparent system, as there are no hidden quirks of domain knowledge enciphered deep within a program's structure. Detailed external validation of the data and calculations can be carried out, ensuring semantic compatibility and computability, and supporting reproducible science. 
We also hope to be able to track semantic relations, so that aspects such as provenance, uncertainty and sensitivity can be threaded through the execution path, and embedded into the final document.

\section{System Overview}
The Declaratron comprises two main components: an XML engine \footnote{based on XML-XOM, \url{http://www.xom.nu}}, which provides macros, resolution, tree manipulations, decoration, validation and specification of computation\footnote{\url{https://bitbucket.org/petermr/declaratron}}; and SCMathML\footnote{\url{https://bitbucket.org/mo_seph/scmathml/wiki/Home}}, a MathML engine written in Scala\footnote{a JVM language which combines functional programming and object orientation, \url{http://scala-lang.org}}, which can evaluate MathML equations using the context provided by the document---see Figure \ref{fig:DecorationProcess} for an overview. Connections are made to domain specific blackboxes (e.g. JUMBO for chemistry); the most common results are typed numeric quantities evaluated by MathML processing and serialized as CML.

\fig[1.0]{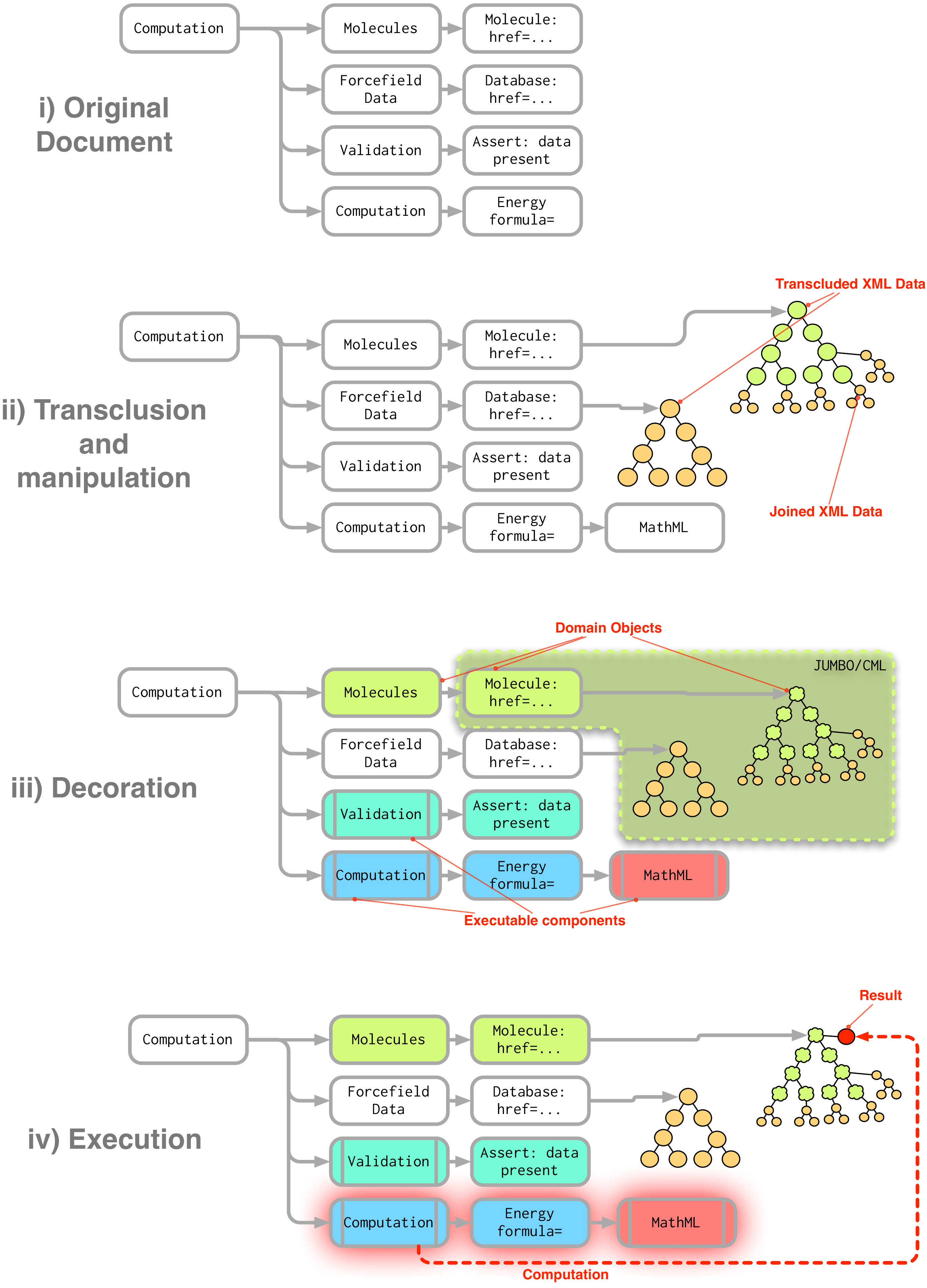}{Overview of system operation. i) original document; ii) manipulated XML document iii) decorated document with executable domain objects iv) executing a computation}

\subsection{Executable MathML}
\begin{figure}[h]
\centering

\begin{subfigure}[b]{0.9\textwidth}
\centering
\begin{lstlisting}
parse(<apply><plus/><cn>2</cn><cn>2</cn></apply>).eval()
\end{lstlisting}
\caption{Simple addition---returns 4}
\label{fig:addition}
\end{subfigure}
\newline

\begin{subfigure}[b]{0.9\textwidth}
\centering
\begin{lstlisting}
parse(<apply><plus/>
    <apply><power/><ci id="x">x</ci><cn>2</cn></apply>
    <ci id="c">c</ci>
</apply>).eval(Context( "x" -> 2, "c" -> 4 ) )
\end{lstlisting}
\caption{Using values provided in a formula}
\label{fig:formula}
\end{subfigure}
\newline
\begin{subfigure}[b]{0.9\textwidth}
\centering
\begin{lstlisting}
parse(<apply><sum/>
    <bvar><ci>atom</ci></bvar>
    <condition><apply><in/> <!-- iterating over the set "atoms" -->
        <ci>atom</ci><ci type="set">atoms</ci>
   </apply></condition>
	<apply> <!-- get value from object -->
        <csymbol func="getMass">w</csymbol>
        <ci>atom</ci>    			
    </apply>
</apply>).eval(Context( "atoms" -> cml:molecule.getAtoms() )
\end{lstlisting}
\caption{Summing atomic masses in a molecule. NOTE: In future versions, getMass will be replaced with a URI, and a dictionary approach will be used to map URIs onto functions in blackbox libraries.}
\label{fig:summation}
\end{subfigure}
\newline
\newline
\newline
\begin{subfigure}[b]{0.9\textwidth}
\centering
\includegraphics[width=0.3\textwidth]{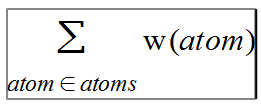}
\caption{Visual rendering of atomic mass summation from Figure \ref{fig:summation}}
\label{fig:summationRender}
\end{subfigure}
\end{figure}

Content MathML (as distinct from Presentation MathML) has a semantic basis---we can have an idea of how links should be made between nodes in a parsed MathML document
and mathematical concepts. A fragment of MathML is not executable on its own, however---it specifies formulae to use, but not what to do with them or how to compute them. Since MathML does not formally define evaluation semantics (although it is tied to OpenMath, and there is a history of evaluating computational algebra) we propose and implement the following:

\begin{enumerate}
\item A MathML fragment can be evaluated, and will return a result of a specific type; Figure \ref{fig:addition}) evaluates $2+2$ and returns $4$.

\item Variables in formulae can be ``bound" to different values; in MathML these are called content identifiers---\lstinline{<ci>}---as distinct from  ``content numeric" (\lstinline{<cn>})--- and return a value by looking for a \lstinline{<bvar>} (“bound variable”) with the same name. Hence, a Context must be provided, mapping from BVars to either values or objects from which values can be obtained. Figure \ref{fig:formula} evaluates $x^2+c$, with $x=2$ and $c=4$, returning $8$.

\item Java and Scala objects can be bound, in order to create lists or sets of values. Additionally, domain specific objects, can then be queried to provide numeric values as necessary. This can (currently) be done by several methods, such as: i) calling named functions on objects (Figure \ref{fig:summation}, second half); ii) running XPath queries to select values---\newline
\texttt{./cml:property/cml:list/cml:scalar[@dictRef='ff:k']}
selects the forcefield spring constant (\texttt{ff:k}) from a list of properties, relative to the current node.
These are relatively ad-hoc techniques, based on evolutionary growth of functionality, and will be replaced with a more formal URI and dictionary approach to mapping semantics onto blackbox objects. 
\item Binding can happen as part of an iteration, for example when summing over a set of values. The first half of Figure \ref{fig:summation} iterates over the atoms in a molecule, binding each one in turn to ``atom'', and then evaluating the code in the second half.
\end{enumerate}

These are all valid MathML expressions---Figure \ref{fig:summationRender} shows a standard rendering of the expression in Figure \ref{fig:summation}.

In order to implement this specification, we construct a parallel tree of Scala objects which can carry out computation\footnote{Arguably, we could have done this by decorating the existing tree, and we may do this in future developments. However, this separation helped to create a MathML engine which was distinct from any particular platform specific XML representation.}. This is constructed of objects which represent simple expressions such as addition and subtraction, complex functions, and iterations over sets, lists and matrices. Each expression is expected to return a value, and values can be typed. This construction is carried out using the Scala Parser Combinators library, to give high level pattern matching (an LL* grammar) with tight code integration. 

\subsection{Semantics} 
Where defined by the MathML specification, mathematical semantics are hardcoded into the Scala MathML engine. The semantics of physical quantities are defined by standoff CML dictionaries (roughly similar architecture to MathML CDs). These indicate human semantics by descriptive text and machine semantics through types (e.g. dimensions of scientific units). The Declaratron XML dialect can be used in dictionaries to indicate computable conversions. In the case of chemistry, many of the operations are hardcoded in the JUMBO framework (e.g. \texttt{bond.getLength()}, \texttt{molecule.getMass()}). Together, these provide maths and chemistry ``blackboxes'' which usually do not have to be recoded for new problems.

\subsection{Declaratron XML and Document Preparation}
So far, we have dealt with two XML dialects---MathML, and to some extent CML---and given indications for how they can be related to each other. In order to operationalise these relationships, and carry out computation, Declaratron XML (DeXML) is used to specify document manipulation operations and computational tasks. The vocabulary used is:
\begin{itemize}
\item \lstinline{<sem:computationalDocument>} is the overall container and organizer;
\item \lstinline{<sem:editor>} allows the document to modify itself using copy, transform, move and delete operations;
\item \lstinline{<sem:assert>} tests components against scalar values or complete (XML) files;
\item \lstinline{@href} allows input of files (transclusion-copy);
\item \lstinline{<sem:writer>} outputs sections of the document;
\item \lstinline{<sem:functionalForm>} specifies a MathML expression which can be bound to other domain semantics;
\item \lstinline{<sem:computation>} evaluates a \lstinline{<sem:functionalForm>} either once or in an algorithm (e.g. an optimization routine).
\end{itemize}

In order to create an executable XML document, a number of steps have to be carried out:
\begin{enumerate}
\item Resolution of symbols---variables which can be defined and used later; 
\item References have to be resolved recursively. Within DeXML, \texttt{href} attributes are used to include content in other files---for example, common formulae, or databases of object properties. Basic provenance is recorded: a) any provenance attached to the transcluded data, and b) the locations from which the data was retrieved (the hrefs).
\item The tree is decorated, by promoting standard XML elements (\texttt{nu.xml.Element}) to computationally active objects, e.g. \texttt{org.xmlcml.cml.element.CMLAtom}. This allows access to domain specific calculation---for example atomic weights or interatomic distances. 
\item Operations can be carried out on the tree, e.g. attaching bond information from a database of bonds---or generally tidying.
\item Tree integrity can be checked, making sure that there is data in the right places or operations over units, checking or translating numerical values.
\end{enumerate}
The XML document is now a computational object with all necessary data.

\subsection{Computation}

When the document is fully decorated, it can be examined to find executable nodes. A Visitor pattern is used, which searches for any executable elements in the tree and then runs them. This execution can include simple calculation, summation, optimisation and so on. The general form of the operation is:
\begin{enumerate}
\item a MathML element is parsed into an executable structure
\item a set of target objects is created from an XPath selector
\item For each target object, the MathML element is given data from the tree, including the target object, and then asked to carry out a calculation. 
\item In its simplest form, this could be appending a single numeric value to an object in the tree---for example, calculating the current energy of a molecule in its initial position. More complex operations are also possible---for example, if a molecule's structure is optimised using the MathML forcefield given, then a copy of the molecule with the new atomic coordinates is added to the document.
\end{enumerate}

The final document is serialised, giving a complete record of the data and equations used,  their sources, the calculations carried out and any intermediate steps. Granular output is also possible by specifying subtrees using XPath, and serialising those objects through the course of the calculation.

\section{Case study: computing forcefield energy}
We have converted the Amber equation given in Figure \ref{fig:amber} to MathML, combined it with a forcefield of several hundred parameters in CML, with the geometry of acetic acid (in CML) and computed the energy. This agrees with the result from the Amber program. In addition we have taken distorted geometries and optimised them using a non-derivative optimiser (which uses a grid of single-point energies to find an optimum). At present we are concerned with correctness of problem description and correctness of result, and not with speed.

The details of this study are given in more detail in an invited chapter for "Implementing Reproducible Computational Research" \cite{murrayrust2013declaratron-book} (draft freely available) where we describe the steps in preparing the Declaratron for the study.

\section{Discussion}

Carrying out the case study gave several insights which have contributed to the language; in particular:
\begin{itemize}
\item Unit testing was utterly essential to developing trust in the system as a whole, and providing support for claims of reproducibility. Through the course of development, we created over a hundred tests for various system properties. This led to the inclusion of \texttt{assert} elements in DeXML, so that as well as blackbox libraries, the operation of the code on actual data can be checked, and readers can be guided through expected outcomes.
\item Many expressions---especially XPath and file locations---become unwieldy and repetitive; it was essential to be able to define variables for common tasks and locations in order to increase the human readability of the document.
\item Many data and formulae are in forms that make semantic computation difficult; a significant, although one-off, effort was needed to translate the AMBER forcefield input (FORTRAN) into structured CML.
\item There is a gradual process of defining higher level semantics which are general, and increase the expressive power of DeXML; while this decreases local explicitness, it allows for greater re-use of code, and human readability. Creating variables is an example of this. 
\item There is a balance between implicit and explicit semantics; in general, explicit declarations are more verbose and cumbersome. As a general principle, we found that we built functionality in an implicit manner to start with, in order to understand the operations necessary, and replaced it with increasingly explicit versions once sufficiently concise representations could be found. As an example, formulae were initially applied to molecules to calculate a single energy value. Over time, this implicit application was converted into a general application of functional forms to data using algorithms, with clearer semantics about what should be done and where the results should go.
\end{itemize}

The use of XML for the complete representation brings many advantages through leveraging existing widespread XML tools and libraries. For example XPath allows very complex searches, such as \texttt{//m:apply[m:log and m:apply[m:sin]] and m:apply[m:log and m:apply[m:cos]]} to retrieve any expression containing a sum including $log(sin(x)$ and $log(cos(y)$. This would allow computations (input, intermediate, or final output) to be searched by mathematical forms. Combined with transclusion of formulae, this can make sharing of computational techniques both easy and automatable. Since MathML can be presented in a human readable form, selecting alternative formulations, or comprehending novel specifications does not require learning XML or a programming language, and existing editing and visualisation tools can be used.

Declaratron objects can be annotated, and act as containers for meta-data as well as computational data. This gives an opportunity for:
\begin{itemize}
\item Maintaining provenance information, by annotating computational or data nodes.
\item Uncertainty analysis (data annotated with uncertainty ranges, or distributions).
\item Fine-grained sensitivity analysis and logging of computation;a MathML node can track the values it produces through the course of execution.
\item Integration with editors, and into the publishing pipeline, to provide full executable papers. If all the data in a paper were open, then it could contain all its own computation, and be runnable by any end user).
\end{itemize}

This last point relates to supporting well tested and reproducible research. We argue that scientific codes should have test Declaratron examples which compute expected results against which the main code can be tested, simultaneously providing demonstrations of correctness and documentation. These can be linked from papers which use computational elements, so that end users can verify the entire results chain of a given paper. Since the Declaratron XML is not implementation specific, alternative implementations could be used. As an example, the Scala MathML engine used here is appropriate for running or testing small computations, but an alternative implementation could produce paralellizable GPGPU\footnote{General Purpose GPU allows code to be run on graphics processors, which can dramatically speed up computation for certain tasks} code so the same formula specification can be used in large simulations. This could be especially relevant for fields where public confidence in science is crucial, e.g. climate science\footnote{http://clearclimatecode.org/goal/}

\subsection{Conclusions}

We have argued that the communication of computation in the current literature is not semantically complete, and can hide domain knowledge, leading to important operational features being buried deep in implementations. We have proposed an approach using executable MathML and standards-compliant XML processing which makes the links between computation and domain objects explicit and transparent. Finally, we have discussed how this can aid sharing of scientific knowledge, metadata integration and reproducible science.

\bibliographystyle{splncs}
\bibliography{main}


\end{document}